\documentclass[aps, prb, preprint, superscriptaddress, longbibliography]{revtex4-1}     
\usepackage{graphicx}
\usepackage{dcolumn}
\usepackage{bm}
\usepackage{natbib}
\usepackage{upgreek}
\usepackage{amsmath}
\usepackage{amssymb}

\newcommand*{\Sm}{SmFe$_3$(BO$_3$)$_4$}

\newcommand*{\Ba}{Ba$_2$CoGe$_2$O$_7$}
\newcommand*{\vect}[1]{\mathbf{#1}}
\newcommand\at[2]{\left.#1\right|_{#2}}

\begin{document}

\title{Controlling of light with electromagnons}

\author{D. Szaller}
\author{A. Shuvaev}
\affiliation{Institute of Solid State Physics, Vienna University of
Technology, 1040 Vienna, Austria}
\author{A. A. Mukhin}
\author{A. M. Kuzmenko}
\affiliation{Prokhorov General Physics Institute, Russian Academy of
Sciences, 119991 Moscow, Russia}

\author{A. Pimenov}
\affiliation{Institute of Solid State Physics, Vienna University of
Technology, 1040 Vienna, Austria}

\begin{abstract}
Magnetoelectric coupling in multiferroic materials opens new routes to control the propagation of light. The new effects arise due to dynamic magnetoelectric susceptibility that cross-couples the electric and magnetic fields of light and modifies the solutions of Maxwell equations in media. In this paper two major effects will be considered in detail: optical activity and asymmetric propagation. In case of optical activity the polarization plane of the input radiation rotates by an angle proportional to the magnetoelectric susceptibility. The asymmetric propagation is a counter-intuitive phenomenon and it represents different transmission coefficients for forward and backward directions. Both effects are especially strong close to resonance frequencies of electromagnons, i.e. excitations in multiferroic materials that reveal simultaneous electric and magnetic character.

\end{abstract}

\date{\today}


 \maketitle

\section{Introduction}

Magnetoelectric coupling in multiferroics leads to appearance of new excitations that carry both, magnetic and electric dipole moments. These excitations are expected due to mixing of magnetic and electric order parameters in these materials. Thus, original excitations - magnons and phonons - are getting "coloured", i.e. magnons can be excited by an ac electric field and the phonons start to carry a magnetic dipole moment.

The existence of magnetoelectric excitations was pointed out already long time ago by Smolenskii and Chupis\cite{smolenskii_ufn_1982}, which was followed by further investigations especially in the Russian literature \cite{krivoruchko_jetp_1988, eremenko_book, turov_jetpl_2001}. The general idea of new excitations is rather simple: as long as the magnetoelectric coupling is absent, the magnons and phonons are independent and show no cross-coupling. The inclusion of the coupling into the equations of motion mixes the eigenmodes of the magnetic and electric subsystems thus leading to magnetoelectric excitations. Unfortunately, in most cases and for the materials investigated at that time the coupling was mainly due to relativistic corrections and therefore rather small. Typical estimates of the order of the effects gave the values e.g. for the magnetoelectrically-induced frequency shifts not more than $\Delta \omega / \omega \sim 0.01$ thus making the experimental observation nearly impossible.

The revival of the magnetoelectric effect~\cite{fiebig_jpd_2005} has led to discovery of new material systems with strong coupling of electricity and magnetism. For example, orthorhombic rare-earth (R) manganites with general formula RMnO$_3$ were materials where unusually strong magnetoelectric excitations have been detected and they were termed electromagnons. After first observation~\cite{pimenov_nphys_2006} in GdMnO$_3$ and TbMnO$_3$ the list of compounds with electromagnons has grown rapidly~\cite{sushkov_prl_2007, takahashi_prl_2008, kida_prb_2008, pimenov_prb_2008, aguilar_prl_2009}, see Refs.~[\onlinecite{tokura_rpp_2014, dong_advph_2015}] for review. Recently, electromagnons with electric dipole contribution as large as $\Delta \varepsilon \sim 40$ have been reported in multiferroic borates~\cite{kuzmenko_prb_2014}.

\begin{figure}[tbp]
\centerline{\includegraphics[width=0.99\linewidth,clip]{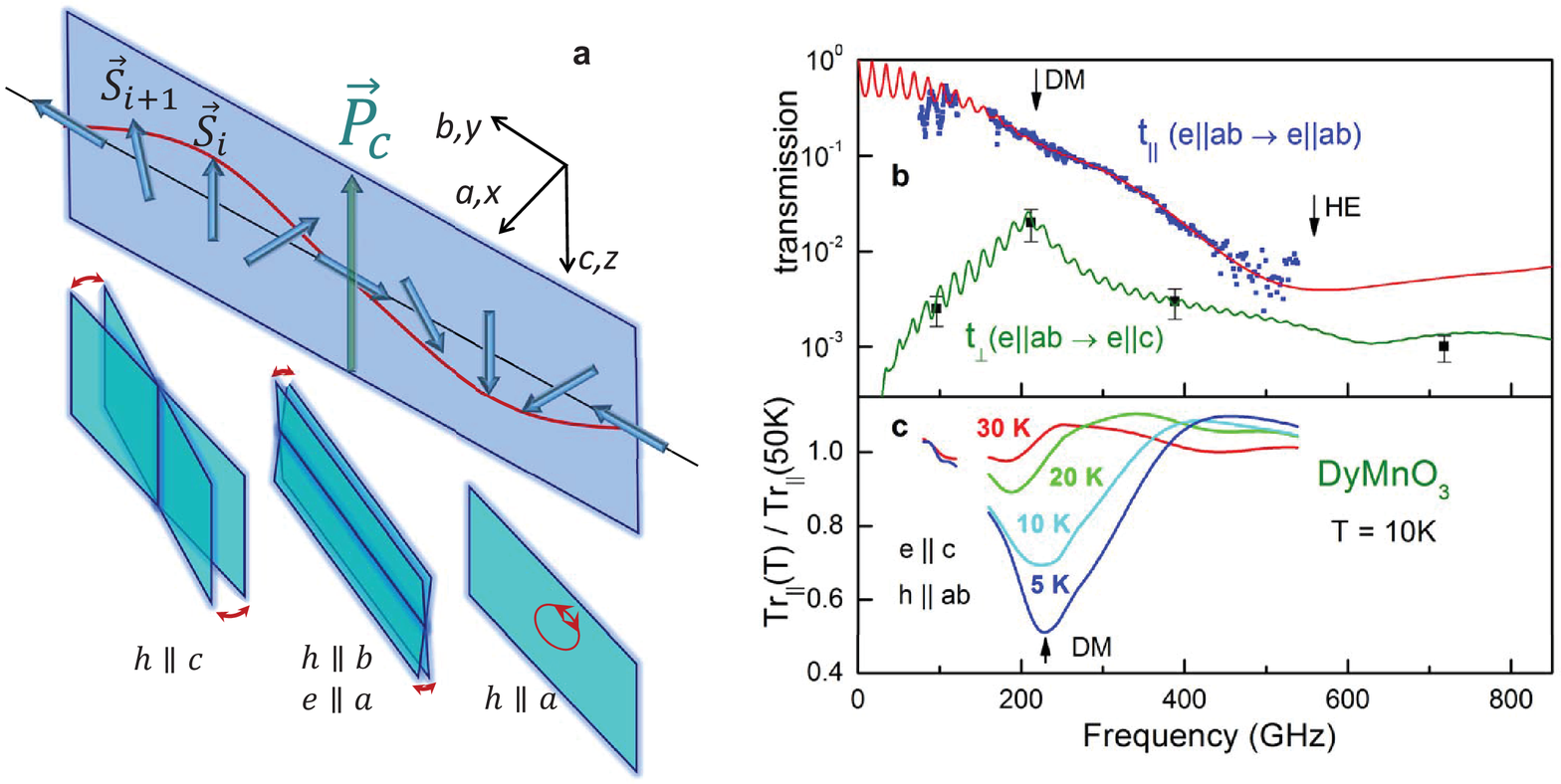}}
\caption{
Illustration of electromagnons in DyMnO$_3$ as example. \textbf{a}. Electromagnon of the ordered magnetic structure that breaks the space inversion symmetry. Shown is an example of cycloidal structure realized in multiferroic manganites. The static polarization and the magnetoelectric character of the electromagnon mode with $h\|b$ and $e\|a$ is due to inverse Dzyaloshinskii-Moriya (DM) coupling~\cite{katsura_prl_2005, mostovoy_prl_2006}. \textbf{b,c}. Simultaneous observation\,\cite{shuvaev_prl_2013} of the DM electromagnon and strong Heisenberg exchange (HE) electromagnon in DMnO$_3$.  Experiment: blue points - transmission in parallel polarizers, black squares - transmission in crossed polarizers that is sensitive to magnetoelectric susceptibility only, solid lines in \textbf{c} - transmission in the geometry sensitive to magnetic channel only ($h\|b$, see panel \textbf{a}). Solid lines in \textbf{b} - model calculations assuming simultaneous presence of both electromagnons. Adapted from Ref.\,[\onlinecite{shuvaev_prl_2013}]
}
\label{f_elmag}
\end{figure}

In several cases, including those described in this paper, the electromagnons originate from the eigenmodes of the ordered magnetic structure. If the magnetic symmetry of the structure allows the magnetoelectric coupling, static electric polarization arises simultaneously with magnetic ordering. One possible configuration is illustrated in Fig.\,\ref{f_elmag}\textbf{a} showing a cycloidal magnetic ordering (see also the paper by T. Kurumaji, DOI: 10.1515/PSR.2019.0016). The cycloidal ordering breaks the inversion symmetry leading to an electric polarization in the form~\cite{katsura_prl_2005, mostovoy_prl_2006}
\begin{equation}
\vect{P}_0 \sim  \vect{e}_{ij} \times (
\vect{S}_i \times \vect{S}_{j}) \ . \label{eq_DM}
\end{equation}
Here $\vect{S}_i$ and $\vect{S}_{j}$ are the neighbor Mn$^{3+}$
spins within the spin cycloid and $\vect{e}_{ij}$
is the vector connecting them (see Fig.~\ref{f_elmag}\textbf{a}).
In Eq.\,(\ref{eq_DM}) the electric polarization is directly coupled to the magnetic order opening the magnetoelectric excitation channel. In this case the magnons can be excited by the electric field of light. As mentioned above, due to the weakness of the Dzyaloshinskii-Moriya (DM) mechanism, these electromagnons in multiferroic manganites are difficult to observe. In Fig.\,\ref{f_elmag}\textbf{b} this excitation is hardly seen in direct transmission (blue points). In order to detect the DM electromagnon in DyMnO$_3$, either magnetoelectric channel (black squares and green line) or purely magnetic channel (Fig.\,\ref{f_elmag}\textbf{c}) can be used. However, in several examples, including \Sm~ discussed below, even the "direct" electromagnon may be of substantial intensity.

Especially in several new multiferroic systems the microscopic mechanism of Heisenberg exchange (HE) striction may lead to strong electromagnons. In this case the coupling is proportional to the scalar product of the neighbour spins $S_i \cdot S_{j}$ and is orders of magnitude stronger than the relativistic DM mechanism. In DyMnO$_3$ (Fig.\,\ref{f_elmag}\textbf{b}) as in several other multiferroic manganites the HE electromagnon is a zone-boundary magnon that gets an electric dipole moment due to simultaneous existence of the DM coupling. Not going into the details of the specific models~\cite{aguilar_prl_2009, miyahara_condmat_2008, mochizuki_prl_2010} we note here that in addition to the exchange striction mechanism a cycloidal magnetic structure is necessary to make the electromagnons observable by optical spectroscopy. Surprisingly, to realize the cycloidal magnetic structure the DM mechanism is essential. Therefore, in some sense the interaction of both mechanisms leads to the dynamic magnetoelectric coupling. In such cases, two types of electromagnons may be expected: strong "Heisenberg" electromagnon and much weaker DM electromagnon. Indeed, in TbMnO$_3$ and DyMnO$_3$ (Fig.\,\ref{f_elmag}) both excitations could be reliably observed and separated in the spectra\cite{aguilar_prl_2009, shuvaev_prl_2010, shuvaev_prl_2013}.

Here we review possible routes to control light using the magnetoelectric excitations. We demonstrate that magnetoelectricity can do much more than only giving electric dipole moment to the magnons. The appearance of magnetoelectric terms in Maxwell equations leads to completely new phenomena like optical activity or asymmetric light propagation that will be considered in the following sections. These effects arise because the magnetoelectic susceptibility opens new ways in controlling of light.
The remaining of this paper is organized as follows. In Section II we connect optical activity with magnetoelectric excitation (II.A) and present two examples of experimental observation of the polarization rotation due to magnetoelectric effect. We review the data on two intensively studied multiferroics \Sm~ (II.B) and \Ba~(II.C), both presenting an easy-plane type antiferromagnetic ground state strongly coupled to the electric polarization. In section III the effect of directional anisotropy due to magnetoelectric effect is discussed, again using two examples, \Sm~ (III.A) and \Ba~(III.B). Sections IV and V give the Summary and Outlook, respectively.

\section{Optical activity with electromagnons \label{s_OA}}

In this Section we describe how the magnetoelectric terms in multiferroics can lead to the polarization rotation of light. As can be already derived from the considerations of the  sum rules~\cite{szaller_prb_2014}, the existence of a static magnetoelectric effect must be accompanied by a nonzero dynamic magnetoelectric susceptibility.

We start with a simple derivation of the connection between the magnetoelectric susceptibility and the optical activity. Within  experimental conditions the optical activity may be measured as the rotation of the polarization plane of light after passing through the sample. We assume that a linearly polarized light is incident on the sample and only the magnetoelectric susceptibility is responsible for the rotation.
Even in the simplified case the calculation of the transmission through the magnetoelectric sample is a difficult task. In real experiments the calculations are done numerically, utilizing e.g. the Berreman approach~\cite{berreman_josa_1972, kuzmenko_prb_2014, shuvaev_sst_2012} or the M\"{u}ller matrix formalism~\cite{stanislavchuk_rsi_2013}. However, in  the approximation of thin sample useful relations between polarization rotation and dynamic magnetoelectric susceptibilities can be obtained as we demonstrate below.

\subsection{Polarization rotation via magnetoelectric effect}

\begin{figure}[tbp]
\centerline{\includegraphics[width=0.4\linewidth,clip]{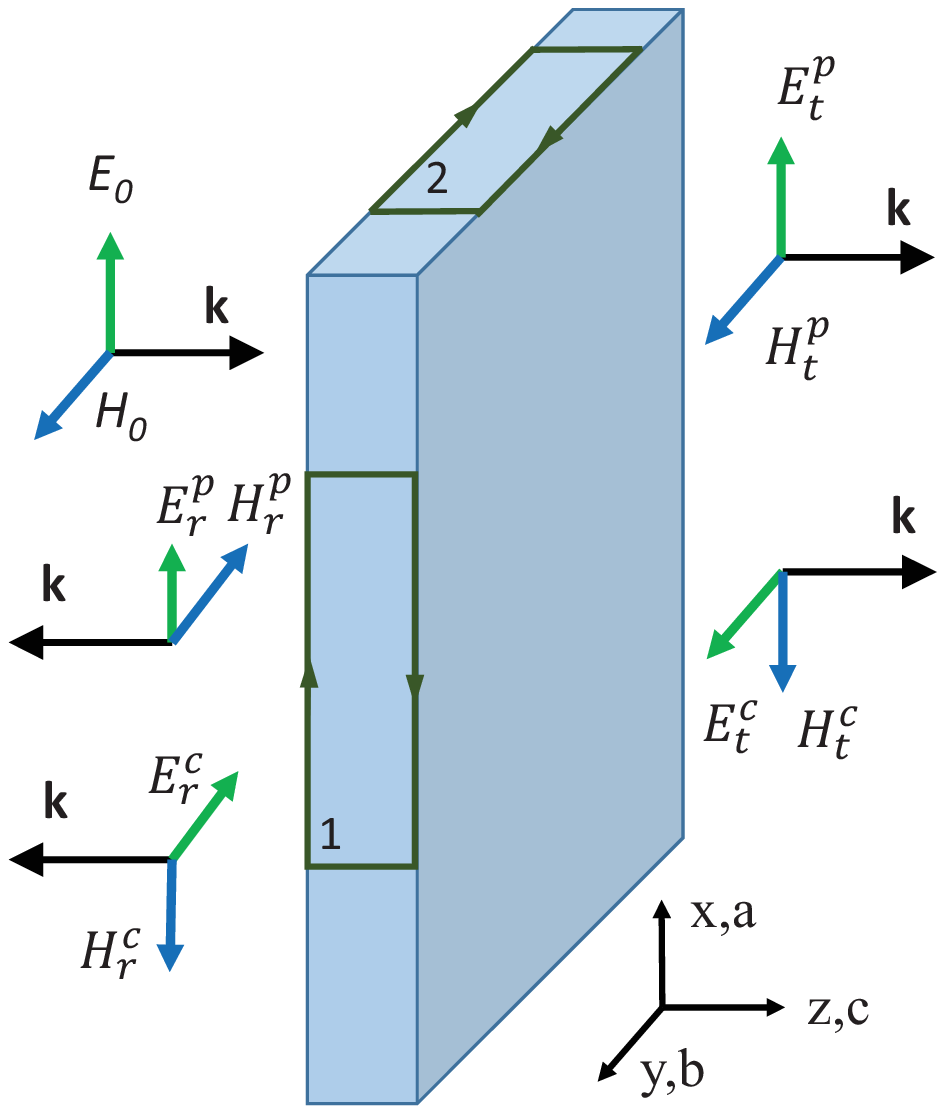}}
\caption{
Geometry of the experiment on polarization rotation of a linearly polarized wave with $E_0 \| x$ in passing through a thin sample. Green arrows $E$ and blue arrows $H$ denote the electric and magnetic fields of a given light beam and the black arrow $\vect{k}$ shows the propagation direction. Subscripts $t$ and $r$ indicate the transmitted and reflected waves, respectively. Superscripts correspond to the components parallel ($p$) and perpendicular ($c$) to the polarization of the incident wave. Integration paths $1$ and $2$ are drawn by dark green.
}
\label{fsch}
\end{figure}

The rotation of the polarization plane as a result of the magnetoelectric susceptibility may be calculated in a simple way in the approximation of thin sample. Within this approximation the plane waves propagating within the sample and the exponential coordinate dependence $~\sim \exp{(\pm ikz)}$ is reduced for $kz \ll 1$ to linear terms only: $\exp{(\pm ikz)} \approx (1 \pm ikz)$. Here $k$ is the wave-vector that results  from the solution of the Maxwell equations within the sample and $z$ is the propagation direction. In addition, we simplify the geometry assuming the plane-parallel sample with the wave propagating perpendicular to the surface and neglecting possible effects of asymmetric propagation (see Sec.~\ref{chdir}). The usefulness of the present approximation is a linear coordinate dependence of all fields within the sample.

We consider the linear polarization state of the incident light with oscillating electric field $E_0 \| x$-axis. After interaction with the sample, the reflected and transmitted waves become elliptical in general case, that can be presented as a sum of two linearly polarized waves with polarizations along the $x$-- and $y$--axes, respectively. The geometry of the problem and the notations are given in Fig.~\ref{fsch}.
In a first step the third Maxwell equation
\begin{equation}\label{eqmaxw}
  \oint \vect{E} d\vect{s} = - \frac{\partial}{\partial t} \iint \vect{B} d\vect{A}
\end{equation}
is applied to the path \textbf{1} shown in Fig.~\ref{fsch}, taking into account the path integration along the sample surface only. Here $\vect{B}$ stands for the magnetic flux density. In this case the left hand side of Eq.~(\ref{eqmaxw}) per unit length is reduced to $-E_0 -E_r^p +E_t^p $. The surface integral on the right hand side assuming the $\sim \exp{(-i\omega t)}$ time dependence can be written as:
\begin{equation*}
  - \frac{\partial}{\partial t} \iint B dA = i \omega d B_y \ .
\end{equation*}
Here $d$ is the sample thickness and $B_y$ is the averaged value of the magnetic field across the sample. In the present approximation it may be rewritten as a half sum of the values at the left and right sample surfaces. This useful simplification will be applied to all components of the fields.

In a similar way, Eq.~(\ref{eqmaxw}) is applied to the path \textbf{2} in Fig.~\ref{fsch} and then the fourth Maxwell equation
\begin{equation*}
  \oint \vect{H} d\vect{s} =  \frac{\partial}{\partial t} \iint \vect{D} d\vect{A}
\end{equation*}
is applied to both paths in Fig.~\ref{fsch}, where $\vect{H}$ stands for the magnetic field and $\vect{D}$ is the electric displacement field.
As a result, the following set of four equations is obtained:
\begin{equation} \label{eqmaxwell}
 \begin{aligned}
  -E_0 - E_r^p + E_t^p &=& i \omega d B_y \\
  -E_r^c - E_t^c &=& i \omega d B_x \\
  H_r^c - H_t^c &=& -i \omega d D_y \\
  H_0 - H_r^p - H_t^p &=& -i \omega d D_x \ .
 \end{aligned}
\end{equation}
The averaged fields $\vect{D}$ and $\vect{B}$ are connected to $\vect{E}$ and $\vect{H}$ inside the sample via the material equations:
\begin{equation}
\begin{array}{c}
 \vect{B} = \mu_0\hat{\mu} \vect{H} + \sqrt{\varepsilon_0 \mu_0}\hat{\chi}^{me} \vect{E} \\
 \vect{D} = \sqrt{\varepsilon_0 \mu_0} \hat{\chi}^{em} \vect{H} + \varepsilon_0 \hat{\varepsilon} \vect{E} \ . \\
\end{array}
\label{eqmater}
\end{equation}
Here $\mu_0$ and $\varepsilon_0$ stand for the vacuum permeability and permittivity, while $\hat{\mu}$, $\hat{\varepsilon}$ denote the relative magnetic permeability and electric permittivity tensors of the material. The $\hat{\chi}^{me}$ and $\hat{\chi}^{em}$ tensors are the magnetoelectric and inverse magnetoelectric susceptibilities of the sample.

Unfortunately, if we take the susceptibilities in general form, the resulting expressions will be still too complicated. In order to simplify the problem we take the example of \Sm. As has been shown by microscopic calculations~\cite{kuzmenko_prb_2014}, the simplified susceptibilities in case of \Sm~ may be written as:
\begin{equation}
\begin{array}{cc}
\hat{\chi}^m(\omega) =
\left( \begin{array}{ccc}
0 & 0 & 0 \\
0 & \chi_{yy}^m & \chi_{yz}^m \\
0 & \chi_{zy}^m & \chi_{xx}^m \\
\end{array} \right) & \hat{\chi}^{me}(\omega) =
\left( \begin{array}{ccc}
0 & 0 & 0 \\
0 & \chi_{yy}^{me} & 0 \\
0 & \chi_{zy}^{me} & 0 \\
\end{array} \right) \\[3em]
\hat{\chi}^{em}(\omega) = \left( \begin{array}{ccc}
0 & 0 & 0 \\
0 & \chi_{yy}^{em} & \chi_{yz}^{em} \\
0 & 0 & 0 \\
\end{array} \right) & \hat{\chi}^e(\omega) = \left( \begin{array}{ccc}
\chi_{xx}^e & 0 & 0 \\
0 & \chi_{yy}^e & 0 \\
0 & 0 & \chi_{zz}^e \\
\end{array} \right) \ , \\
\end{array}
\label{eqsuscept}
\end{equation}
Here $\hat{\chi}^m$ and $\hat{\chi}^e$ are the magnetic and electric susceptibilities of the material. Compared to Ref.~[\onlinecite{kuzmenko_prb_2015}] we neglect here the terms which are not resonant at the electromagnon frequency.

Because the electromagnetic waves outside the sample are transversal, the $z$ components $B_z$ and $D_z$ are zero. Applying the boundary conditions of the normal components of the fields, $D_z = const$, $B_z = const$ and taking into account the linearity within the sample, we see that $B_z$ and $D_z$ are zero everywhere. This allows us to exclude the $z$-components of the $\vect{H}$ and $\vect{E}$ fields from material equations, Eqs.~(\ref{eqmater},\ref{eqsuscept}). As a result, four "effective" boundary conditions for the transverse components of the fields inside the sample are obtained.

The remaining four equations are necessary to connect the fields across the sample. They are obtained taking into account that the transversal components of $\vect{H}$ and $\vect{E}$ are continuous at the sample surfaces. Again, the fields inside the sample are calculated within the present approximation as an average of the fields at two surfaces. The resulting equations can be written as:
\begin{equation} \label{eqbound}
 \begin{aligned}
  E_0 + E_r^p + E_t^p &=& 2E_x  \\
  E_t^c - E_r^c &=& 2 E_y  \\
  -H_r^c - H_t^c &=& 2 H_x  \\
  H_0 - H_r^p + H_t^p &=& 2 H_y \ .
 \end{aligned}
\end{equation}
Finally, the electric and magnetic fields for the plane waves in vacuum are connected via $H_{r,t}^{c,p}=E_{r,t}^{c,p} / Z_0$ where $Z_0$ is the impedance of the free apace.

Excluding the $\vect{E}, \vect{H}, \vect{D}, \vect{B}$ fields inside the sample from the obtained array of 12 equations the following solution for the polarization rotation ($\theta$) and ellipticity ($\eta$) of a sample with $d$ thickness  is easily calculated for the linear incident polarization $E_0 \| x$:
\begin{equation}
tan(\theta+i\eta) = E^c_t/E^p_t \approx  - \frac{i \omega d}{2}{\tilde{\chi}_{yy}^{em}}
\label{eqtheta}
\end{equation}
The exact solution within the present approximation is given by:
\begin{equation}
  tan(\theta+i\eta) =   E^c_t/E_0 \approx t_c = - \frac{i \omega d}{2}\frac{\tilde{\chi}_{yy}^{em}}{(1-\frac{i \omega d \hat{\mu}_{yy}}{2Z_0})(1+\frac{i \omega d Z_0 \tilde{\varepsilon}_{y}}{2})}
\label{eqtc}
\end{equation}
\begin{equation}
t_p = E^p_t/E_0 = \frac{1}{2} \left[\frac{1+i\omega dZ_0 {\varepsilon}_x /2}{1-i\omega dZ_0 {\varepsilon}_x /2} + \frac{1+\frac{1+i\omega d \hat{\mu}_{yy}}{2Z_0}}{1-\frac{1+i\omega d \hat{\mu}_{yy}}{2Z_0}}\right] \approx 1
\label{eqtp}
\end{equation}
The notations in Eqs.~(\ref{eqtc},\ref{eqtp}) are:
\begin{eqnarray*}
  \tilde{\chi}_{yy}^{em} &=& \chi_{yy}^{em}-\frac{\chi_{yz}^{em} \chi_{zy}^{m}}{1+\chi_{zz}^m} \ , \\
  \hat{\mu}_{yy} &=& \tilde{\mu}_{yy} +\frac{\tilde{\chi}_{yy}^{me} \tilde{\chi}_{yy}^{em}}{(\frac{2}{i\omega d Z_0}-\tilde{\varepsilon}_y)} \approx \tilde{\mu}_{yy} \approx 1 \ , \\
  \tilde{\mu}_{yy} &=& 1+ \chi_{yy}^m - \frac{\chi_{yz}^{m} \chi_{zy}^{m}}{1+\chi_{zz}^m} \approx 1 \ , \\
  \tilde{\varepsilon}_y &=& 1+\chi_{yy}^e - \frac{\chi_{yz}^{em} \chi_{zy}^{me}}{1+\chi_{zz}^m} \ ,\ \mathrm{and} \\
\varepsilon_x &=& \varepsilon_{x,\infty}+ \chi^e_{xx}
\end{eqnarray*}
Here, $\varepsilon_{x,\infty}$ is the high-frequency permittivity, and $t_p$, $t_c$ stand for the transmission in the parallel and perpendicular polarization, respectively. Equation (\ref{eqtheta}) demonstrates that the diagonal term of the magnetoelectric susceptibility in a reasonable approximation is responsible for the polarization rotation.

On a similar manner, the polarization rotation in the perpendicular geometry $E_0 \| y$ can be obtained and up to the leading term it is given by:

\begin{equation}
tan(\theta+i\eta) = E^c_t/E^p_t \approx  - \frac{i \omega d}{2}{\tilde{\chi}_{yy}^{me}}
\label{eqthetay}
\end{equation}
with
\begin{eqnarray*}
  \tilde{\chi}_{yy}^{me} &=& \chi_{yy}^{me}-\frac{\chi_{zy}^{me} \chi_{yz}^{m}}{1+\chi_{zz}^m}
\end{eqnarray*}
being the renormalized magnetoelectric susceptibility.

We note that the physical mechanism of the thin sample approximation corresponds to the interaction of the electromagnetic waves with two sample surfaces. Therefore, strictly speaking, in Eqs.~(\ref{eqtheta},\ref{eqthetay}) the effect of the sample volume is underestimated. In the opposite approximation\cite{mukhin_2018} one neglects the interaction with surfaces and considers the propagation in the bulk only. In this case the polarization rotation is again proportional to the diagonal term of the magnetoelectric susceptibility ${\chi}_{yy}^{me,em}$  with an additional $\pi/2$ phase shift compared to Eqs.~(\ref{eqtheta},\ref{eqthetay}).

\subsection{Optical activity in samarium ferroborate \label{chopt}}

\begin{figure}[tbp]
\begin{center}
\includegraphics[width=1.0\linewidth, clip]{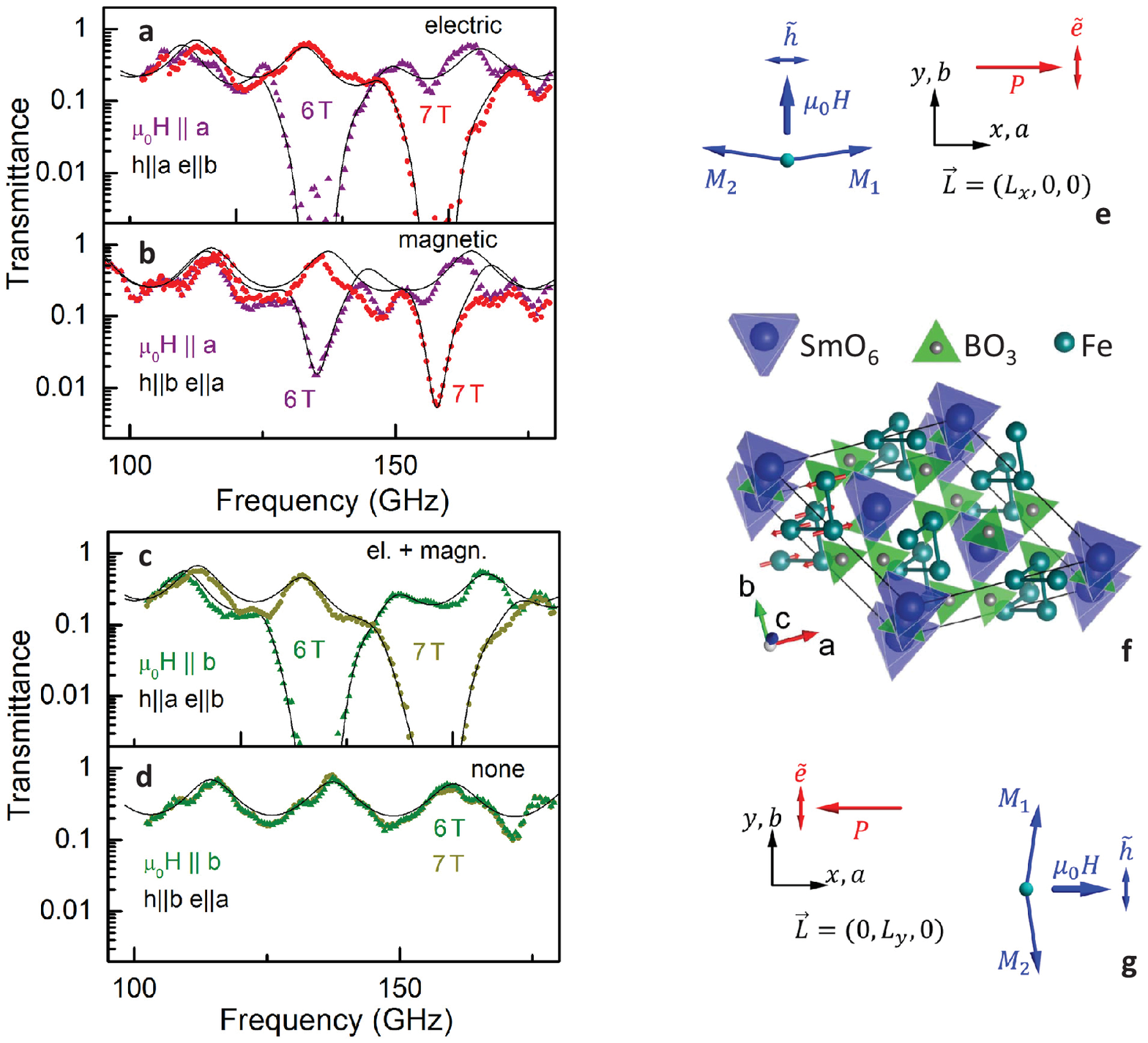}
\end{center}
\caption{Magnetic structure and the excitation of electromagnon in \Sm~ in terahertz spectra~\cite{kuzmenko_prb_2014}. \textbf{a,b} - transmittance spectra for
external magnetic field $\mu_0 \vect{H} \|a $, where the electromagnon can be selectively excited either by electric (\textbf{a}) or by magnetic (\textbf{b}) field of light. \textbf{c,d} - Geometry with magnetic field $\mu_0 \vect{H} \|b $ where the electromagnon is either excited simultaneously in both channels (\textbf{c}) or is silent (\textbf{d}).
The orientation of two iron sublattices in external magnetic fields ($\vect{M}_1$ and $\vect{M}_2$) is shown schematically for two relevant geometries: $\mu_0 \vect{H} \|a $ (\textbf{e}) and $\mu_0 \vect{H} \|b $ (\textbf{g}). The excitation conditions for the electromagnon are indicated by $\tilde{e}$ (electric field) and $\tilde{h}$ (magnetic field). Panel \textbf{f} shows the crystal structure of the magnetoelectric borates.
} \label{fSmrot}
\end{figure}

The mechanisms described in the previous subsection can be demonstrated experimentally in \Sm. As shown in Fig.\,\ref{fSmrot}, samarium ferroborate contains two interacting localized magnetic subsystems given by
Sm$^{3+}$ and Fe$^{3+}$ ions. The iron subsystem orders
antiferromagnetically below $T_N=34~ K$ with an easy-plane magnetic
structure oriented perpendicularly to the trigonal c-axis. Although
the Sm$^{3+}$ moments play an important role in the magnetoelectric
properties of \Sm, they do not order up to the lowest
temperatures. \Sm~ has a non-centrosymmetric trigonal structure with R32 space group \cite{vasiliev_ltp_2006}.

Static electric polarization in multiferroic ferroborates can be
explained by symmetry arguments and by taking into account that
Fe$^{3+}$ moments are oriented antiferromagnetically within the
crystallographic
$ab$-plane~\cite{zvezdin_jetpl_2005,zvezdin_jetpl_2006}. The terms governing the ferroelectric
polarization ($\vect{P}$) along the $a$ and $b$-axis (or $x$ and $y$-axis) are given by
\begin{equation}\label{eqpol}
P_x \sim L_x^2-L_y^2 ,\quad  P_y \sim -2L_xL_y \ .
\end{equation}
Here $\vect{L}=\vect{M}_1-\vect{M}_2$ is the antiferromagnetic vector with $\vect{M}_1$ and
$\vect{M}_2$ being the magnetic moments of two magnetic
Fe$^{3+}$ sublattices. More details of the symmetry analysis of the
static magnetoelectric effects in \Sm~ can be found in Refs.\,
[\onlinecite{zvezdin_jetpl_2005, zvezdin_jetpl_2006, popov_prb_2013}].

Equation~(\ref{eqpol}) allows us to understand the behavior of the
static and dynamic properties in external magnetic fields.  As shown
in Fig.~\ref{fSmrot}\textbf{e}, static magnetic field along the $y$-axis
stabilizes the magnetic configuration with $L_y=0$ and $L_x\neq 0$.
In agreement with Eq.~(\ref{eqpol}) in this case the static polarization
is oriented parallel to the a-axis (Fig.\,\ref{fSmrot}\textbf{g}). For magnetic fields along the
$x$-axis and above the spin flop value $L_x=0$ and $L_y\neq 0$ which
leads to antiparallel orientation of electric polarization with
respect to the a-axis.
\begin{figure}[tbp]
\begin{center}
\includegraphics[width=0.85\linewidth, clip]{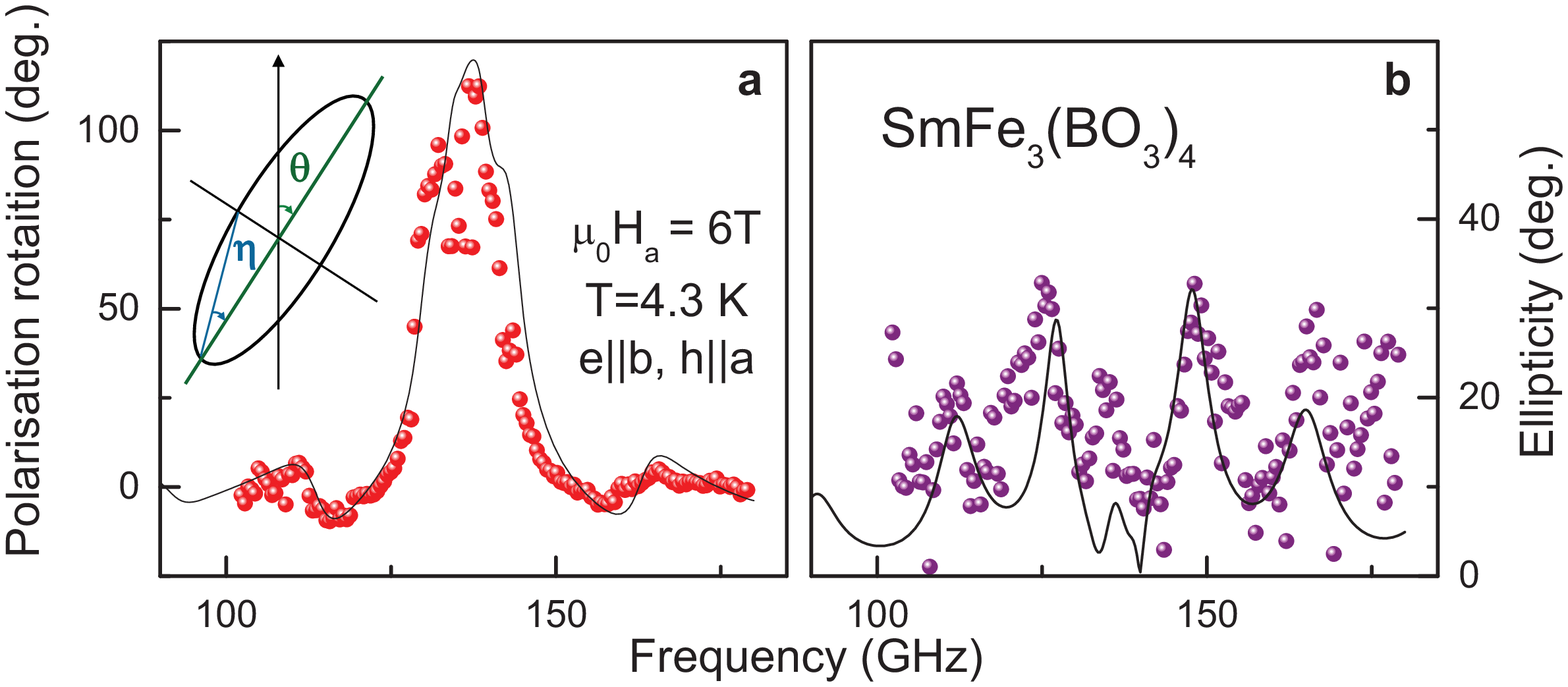}
\end{center}
\caption{Optical activity with electromagnon in \Sm~(Ref~[\onlinecite{kuzmenko_prb_2014}]).
\textbf{a} - Polarization rotation ($\theta$). The inset shows the
definition of rotation angle and ellipticity. \textbf{b} -
Ellipticity ($\eta$). Symbols -
experiment, solid line - model calculations. } \label{fSmrot2}
\end{figure}

The magnetic and
electric excitation channels are connected in \Sm\, because of direct coupling of
antiferromagnetism and ferroelectricity. The low-frequency
magnetoelectric mode of interest corresponds to oscillations of
antiferromagnetic moment $\vect{L}$ in the easy $ab$-plane\cite{kuzmenko_jetpl_2011}. It can be
excited either by an ac electric field with $\vect{e} \bot \vect{P}$ or by an ac
magnetic field $\vect{h} \bot \vect{M}$ (see Fig.~\ref{fSmrot}\textbf{e,g}). Here $\vect{P}$ is the static electric
polarization and $\vect{M}\| \mu_0 \vect{H}$ is a weak field-induced ferromagnetic
moment with $\vect{M} \bot \vect{L}$. Because electric polarization is directly
coupled to the antiferromagnetic order, it becomes possible to
excite the spin oscillations not only by an ac magnetic field but also by
an alternating electric field as well. The excitation conditions of
the electromagnon strongly differ for two magnetic configurations
given in Fig.~\ref{fSmrot}. In particular, the excitation
conditions of the electromagnon for the configuration in
Fig.~\ref{fSmrot}\textbf{a} are given by $\vect{e}\|b$ and for Fig.~\ref{fSmrot}\textbf{b} by  $\vect{h}\|b$, respectively. Therefore, for an
ab-cut of \Sm~  one can selectively excite either electric
($\vect{e}\|b$) or magnetic ($\vect{h}\|b$) component of the electromagnon simply
by rotating the polarization plane of the incident radiation. However, we note that in agreement with Eqs.~(\ref{eqtheta},\ref{eqthetay}) the polarization rotation angle will be closely similar in both geometries.

The excitation conditions for the static magnetic field parallel to the $b$-axis are shown in
Fig.~\ref{fSmrot}\textbf{g} and the corresponding spectra are given in Fig.\,\ref{fSmrot}\textbf{c,d}. In this geometry the conditions of  directional anisotropy are realised. This case will be discussed in Sec.~\ref{DA_Sm}.

Full spectral and polarization analysis of the data in Fig.~\ref{fSmrot}\textbf{a,b} allows us to obtain the angle of the polarization rotation in these experiments~\cite{kuzmenko_prb_2014}. The results are shown
in Fig. \ref{fSmrot2}. In these
experiments a polarization rotation angle exceeding 120 degrees
is obtained for a sample with thickness of $~\sim 2$mm. We
stress out that in agreement with Eqs.~(\ref{eqtheta},\ref{eqthetay}) this rotation arises purely from dynamic magnetoelectric
susceptibilities $\chi_{yy}^{me,em}$, Eq.\,(\ref{eqsuscept}) which are intrinsic for the electromagnon in \Sm.

\subsection{Optical activity in Ba$_2$CoGe$_2$O$_7$}

\begin{figure}[tbp]
\begin{center}
\includegraphics[width=0.99\linewidth, clip]{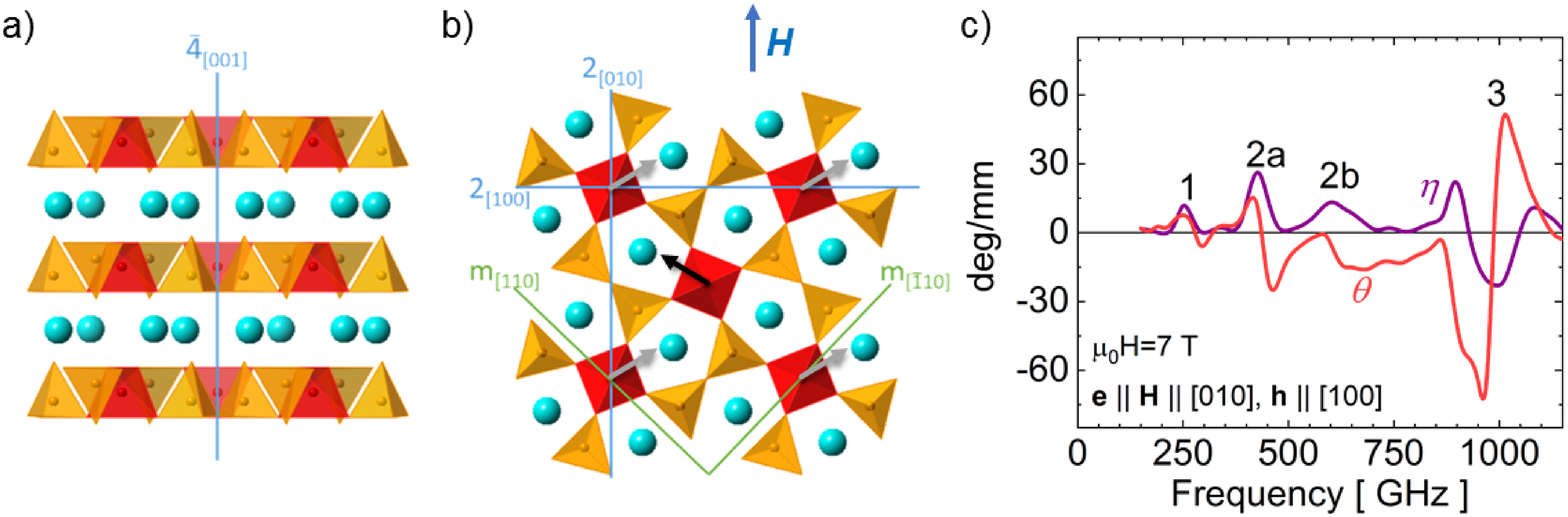}
\end{center}
\caption{$\textrm{P}\overline{4}2_1\textrm{m}$ crystal structure and natural optical activity
of \Ba~ as reproduced from Ref.~[\onlinecite{bordacs_nphys_2012}]. Co$^{2+}$ ions are surrounded
by red tetrahedral oxygen cages, GeO$_4$ tetrahedra are orange and Ba$^{2+}$ spacer ions are shown
by blue spheres. At the corners of the tetrahedral cages oxygen atoms are omitted for clarity.
(a) $3\textrm{D}$ image of the layered crystal structure viewed from the $[100]$ direction,
(b) the tetragonal $(001)$ magnetic easy plane of the crystal. Symmetries of the crystal
structure are presented in (b). Blue lines indicate rotation symmetries around the line,
while green lines are projections of mirror planes. Subscripts are used to label the axes of
rotations and the mirror plane normals. Gray and black arrows represent the chessboard-like
canted antiferromagnetic magnetic ordering in magnetic field (blue arrow) along $[010]$ at $T<T_N$.
(c) The optical rotation $\theta$ (red), and ellipticity $\eta$ (purple) spectra
in $\mu_0H_{[010]}=7\textrm{~T}$ magnetic field\cite{bordacs_nphys_2012}, with incident
polarization $\vect{e}\parallel [010],~\vect{h}\parallel [100]$. The signal of the electromagnon
excitations is marked with numbers: 1 indicates the quasi-ferromagnetic, 2a and 2b the split
quasi-antiferromagnetic, and 3 the spin-streching\cite{penc_prl_2012} mode.
} \label{fBarot}
\end{figure}

In this section we present another example of polarization rotation by electromagnons in Ba$_2$CoGe$_2$O$_7$.
\Ba~, as shown in Fig.~\ref{fBarot}\textbf{a,b}, crystallizes in the non-centrosymmetric, tetragonal
$\textrm{P}\overline{4}2_1\textrm{m}$ symmetry \cite{zheludev_prb_2003,hutanu_prb_2011} with square-lattice
layers of the magnetic Co$^{2+}$ cations perpendicular to the $[001]$ (tetragonal) axis of the crystal.
The nearest neighbours are connected along the $[110]$ and $[1\overline{1}0]$ directions.
The Co$^{2+}$ ions are surrounded by tetrahedral oxygen cages,
which are bridged by GeO$_4$ tetrahedra, while Ba$^{2+}$ spacer ions separate the layers.
The $\textrm{P}\overline{4}2_1\textrm{m}$ space group has $2_1$ two-fold screw axes along
the $[100]$ and $[010]$ directions, while the $(110)$ and $(1\overline{1}0)$ planes correspond to the $\textrm{m}$ mirror planes.

The tetrahedral local symmetry of the oxygen cage results in the quenching of the orbital moment
and thus in the  $S=3/2$ spin-only moment of the Co$^{2+}$ ion.
At low temperatures ($T<T_N\approx 7\textrm{ K}$), the magnetic moments order antiferromagnetically.
The two-sublattice antiferromagnetic state shows an easy-plane character, where spins lie within
the $(001)$ plane, due to the single-ion anisotropy of the Co sites.\cite{zheludev_prb_2003}

The magnetoelectric behavior of \Ba~ in the static limit has been extensively investigated both
experimentally,\cite{murakawa_prl_2010,murakawa_prb_2012, yi_apl_2008} and
theoretically,\cite{yamauchi_prb_2011, perez_mato_acsa_2011, toledano_prb_2011}.
The observed magnetoelectric properties can be understood in the frame of the spin-dependent
metal-ligand $p-d$ hybridization model.\cite{arima_jpsj_2007} Namely, the hybridization of
the $2p$ orbitals of the surrounding oxygen cage with the Co$^{2+}$ $3d$ orbitals depends
on the direction of the Co magnetic moment,\cite{murakawa_prl_2010,murakawa_prb_2012} and
the emergent onsite polarization can be described\cite{murakawa_prl_2010} as

\begin{equation}
\vect{P} \sim \sum_{i=1}^4 \left( \vect{S}\vect{\delta_i}\right)^2 \vect{\delta_i},
\label{pd_equation}
\end{equation}
where the contributions of the four Co-O bonds are summed up. Here,
$\vect{\delta_i}$ vectors stand for the unit vectors pointing from the
Co$^{2+}$ ion towards each neighbouring oxygen. In a basic picture, the more
parallel a given bond to the Co moment is, the stronger the bond gets. Thus
the Co$^{2+}$ ion shifts towards that oxygen atom, generating a local
electric polarization. As evident from Eq.~(\ref{pd_equation}), $\vect{P}$ is
independent of the sign of spins $\vect{S}$, as should be expected based on
time-reversal invariance of the electric polarization.



In moderate external magnetic fields, when the magnetization can be
considered to be proportional to the applied field, within a classical spin
model, the orientation of the sublattice magnetizations can be easily given
as function of the magnitude and direction of the applied magnetic field.
Within this approximation, the magnetically induced electric polarization can
be determined using the spin-dependent hybridization
model\cite{arima_jpsj_2007,murakawa_prl_2010}:
\begin{eqnarray}
P_{[100]} & = & A_{\perp}\left(h_{\perp}-\sqrt{1-h_{\perp}^2}\right)
h_{z}\sqrt{1-h_{z}^2}\sin\phi\label{Pa},\\
P_{[010]} & = & A_{\perp}\left(h_{\perp}-\sqrt{1-h_{\perp}^2}\right)
h_{z}\sqrt{1-h_{z}^2}\cos\phi\label{Pb},\\
P_{[001]} & = &
A_{z}\left(h_{\perp}^2-h_{\perp}\sqrt{1-h_{\perp}^2}-\frac{1}{2}\right)\left(1-h_{z}^2\right)\sin(2\phi)\label{Pc}.
\end{eqnarray}
Here $h_{\perp}=H\sin\theta/H^{Sat}_{\perp}$ and $h_{z}=H\cos\theta/H^{Sat}_{z}$ represent the magnetic field ($H$) relative to the corresponding saturation fields in the tetragonal plane ($H^{Sat}_{\perp}$) and along the tetragonal axis  ($H^{Sat}_{z}$), respectively. Azimuthal ($\theta$) and polar ($\phi$) angles of the field are measured from the $[001]$
and $[100]$ axes, respectively. The phenomenological coefficients $A_{z}$ and $A_{\perp}$ scale the magnetoelectric
coupling for polarization perpendicular and along the tetragonal axis, respectively.

Magnetic field applied along the $[100]$ or $[010]$ axis breaks the
mirror-plane symmetries of the crystal, and the magnetic point group symmetry
decreases to $2_{[100]} 2_{[010]}^\prime 2_{[001]}^\prime $ and
$2_{[100]}^\prime 2_{[010]} 2_{[001]}^\prime $, respectively. Thus, in the
non-chiral \Ba~ crystal, the chirality can be induced by magnetic field
pointing in appropriate directions~\cite{bordacs_nphys_2012}. Opposed to the
chemical chirality of molecules originating from the chiral distribution of
static charges, in our case the specific pattern of the magnetic moments on
the non-centrosymmetric lattice cause the breaking of mirror symmetries.
 The molecular chirality is encoded into the arrangement of atoms, thus the handedness of the molecule cannot be reversed. In contrast, by turning the magnetic field from $[100]$ to $[010]$ the handedness of \Ba~ can be switched, since the ground states of the two cases are related by the $m_{[110]}$ mirror-plane symmetry.\cite{bordacs_nphys_2012}

The chirality of \Ba~ can be proven by measuring the natural optical
activity. Indeed, considerable rotation of the incoming linear polarization
state of light was observed, as reproduced from
Ref.\,[\onlinecite{bordacs_nphys_2012}] in Fig.~\ref{fBarot}\textbf{c}. The
sign of the optical rotation and ellipticity does not change when reversing
the magnetic field, while they change sign for a rotation of the field from
$[100]$ to $[010]$, which operation changes the handedness of chirality. In
the intermediate orientation of the field, when the mirror plane is restored,
the effect disappears.\cite{bordacs_nphys_2012}

Considering the experimental geometry in Fig.~\ref{fBarot}\textbf{c}, the
observed rotation means an induced oscillating electric polarization along
$[100]$ by an oscillating magnetic field along the same direction. However,
based on Eq.~(\ref{Pa}), no electric polarization is expected when the
magnetic moments lie in the $(001)$ plane, giving
$\chi^{me}_{[100][100]}(\omega=0)=0$ in the static limit. The effect can only
persist in the dynamic range, corresponding to a time-reversal symmetric
$\chi^{me}_{[100][100]}(\omega)$.\cite{bordacs_nphys_2012}

In contrast to to the previously discussed case of \Sm, where the optical
activity could be connected to a change in the polarization in the static
limit, the optical activity of \Ba~ has no overall static contribution in the
studied geometry. This is also evident from the derivative-like asymmetric
line shape of the optical rotation angle $\theta$ in
Fig.~\ref{fBarot}\textbf{c}, while for \Sm~ the line shape is symmetric. From
the symmetry point of view, the effect is symmetric under time reversal for
\Ba, while for \Sm~ it is antisymmetric, forbidding/allowing an antisymmetric
static contribution, respectively.

\section{Directional anisotropy via the magnetoelectric coupling \label{chdir}}

Different propagation/absorption of particles or waves along opposite
directions, termed as directional anisotropy, is a strikingly
counterintuitive phenomenon. This effect is non-reciprocal, i.e. it is not
symmetric with respect to the interchange of the source and detector.
Specially in electrodynamics, Helmholtz's reciprocity
theorem\cite{stokes_2009, helmholtz_book_1867, born_book} is a fundamental
principle of geometrical optics, stating that counter-propagating light beams
follow identical paths due to the equal refractive indices of the opposite
directions. However, in that case the medium is implicitly assumed to be
invariant under time and space reversal~\cite{kuzmenko_prb_2019}. When
violating these fundamental symmetries, propagation along opposite directions
can be different. Although the spatial symmetry can be reduced by an external
electric field and the time-reversal symmetry can be broken by an external
magnetic field, the experiments show that internal fields, e.g.
non-centrosymmetric charge order (electric polarization or chiral structure)
combined with magnetization are more effective in producing directional
anisotropy. Namely, the relative difference in the studied quantity
(absorption, phase shift, emitted or diffracted intensity) between forward
and backward propagation is only in the order of $10^{-8}-10^{-2}$ for
centrosymmetric, non-magnetic media placed in external electric and magnetic
fields. However, the spontaneous symmetry breaking magnetism and electric
polarization or chirality of multiferroics can result in relative
non-reciprocal effects reaching the order of
unity.\cite{bordacs_nphys_2012,kezsmarki_nc_2014,kuzmenko_prb_2015} This
optical non-reciprocity, when appearing in the absorption, is termed as
directional dichroism, while in the phase shift it is known as directional
birefringence.

Besides the ground state, strong directional anisotropy in the optical properties also imposes certain conditions on the excitation spectrum. Namely, in the lowest-order process directional anisotropy is produced by the magnetoelectric coupling, requiring simultaneously magnetic-- and electric--dipole resonances. Optical directional anisotropy can be viewed as the constructive/destructive addition of oscillating electric polarization components generated by the electric and magnetic fields of light, where the sign of the interference is opposite for the forward/backward directions. This means that the dimensionless magnetic and electric oscillator strengths of the magnetoelectric excitations must be equal
to the full suppression of the transmission in one direction and leaving the sample transparent for the opposite light propagation.\cite{kezsmarki_nc_2014} Considering that electric excitations are usually by orders of magnitude stronger than magnetic ones, the ideal candidate to observe one-way transparency is an originally magnetic excitation dressed with an electric character, for instance the both magnetic- and electric dipole active electromagnons in multiferroic crystals.

Static magnetoelectric coupling and directional dichroism are closely related, and their connection can be mathematically formulated by using the Kramers-Kronig\cite{kronig_josa_1926,kramers_acif_1927} relations in the static limit\cite{szaller_prb_2014}:
\begin{equation}
\begin{split}
\chi^{me}_{\gamma\delta}(0)=\frac{c}{2\pi}\mathcal{P}\int^{\infty}_{0}{\frac{\Delta\alpha(\omega)}{\omega^{2}}\textrm{d}\omega}.
\label{KhiME_DC}
\end{split}
\end{equation}
Based on the resulting sum rule, the off-diagonal zero-frequency
magnetoelectric properties are predominantly generated by the directional
dichroism of low-frequency excitations, such as electromagnons, which is due
to enhancement of the contribution of absorption difference, $\Delta\alpha$,
at low frequencies via the $\omega^{2}$ denominator. In the sum rule $c$
denotes the speed of light in vacuum, while $\mathcal{P}$ is the Cauchy
principal value integral.

Considering the symmetry of the experimental geometry, all realizations of
the directional dichroism up to now belong to two classes, which are termed
as optical magnetoelectric effect (OME)\cite{arima_jpcm_2008} or toroidal
dichroism\cite{bordacs_prb_2015} and magneto-chiral dichroism
(MChD)\cite{rikken_nature_1997}. However, rigorous symmetry
analysis\cite{szaller_prb_2013} allows other possibilities as well, which
have not been observed so far.

In the case of OME the light beam propagates parallel/antiparallel to the vector product of the polarization and the magnetization, which defines the $\vect{T}=\vect{P}\times \vect{M}$ toroidal moment~\cite{fiebig_jpd_2005}. Since $\vect{T}$ is a well-defined static quantity which has identical symmetry properties to the $\vect{k}$ wave vector of the light beam, the system can distinguish between propagation along/opposite to $\vect{T}$. In other words, none of the symmetries of the system can connect $\vect{+k}$ with $\vect{-k}$, because that symmetry would also annihilate $\vect{T}$. $\vect{T}$ can also be defined locally, i.e. it can be finite without macroscopic magnetization, thus the OME can be utilized to distinguish between antiferromagnetic domains of time-reversed magnetic and identical polar structure\cite{kocsis_prl_2018}. We will present the one-way transparency by OME on the example of the electromagnon in \Sm~in Sec.~\ref{DA_Sm}. Interestingly, due to its strong magnetoelectric coupling, \Sm~also offers another way to control absorption of light via the magnetoelectric effect. In the static limit, by applying electric field, the population of magnetic domains can be tuned and through the selection rules of purely magnetic excitations the optical response can be changed drastically.\cite{kuzmenko_prl_2018}

MChD can arise in chiral media when light travels parallel/antiparallel to the magnetization. With the use of the same argument as above, one can see that none of the symmetries can connect $\vect{+k}$ with $\vect{-k}$, since in chiral systems, i.e. systems having only rotational symmetries, the $\vect{M}$ and $\vect{k}$ vectors have the same transformation properties. The one-way transparency by MChD will be presented in Sec.~\ref{DA_BCGO} on the example of the multiferroic melilite, \Ba.

Finally, directional anisotropy in the microwave range due to skyrmions in multiferroics is discussed by M. Mochizuki, DOI:10.1515/PSR.2019.0017.


\subsection{One-way transparency by optical magnetoelectric effect in samarium ferroborate} \label{DA_Sm}

\begin{figure}[tbp]
\begin{center}
\includegraphics[width=0.85\linewidth, clip]{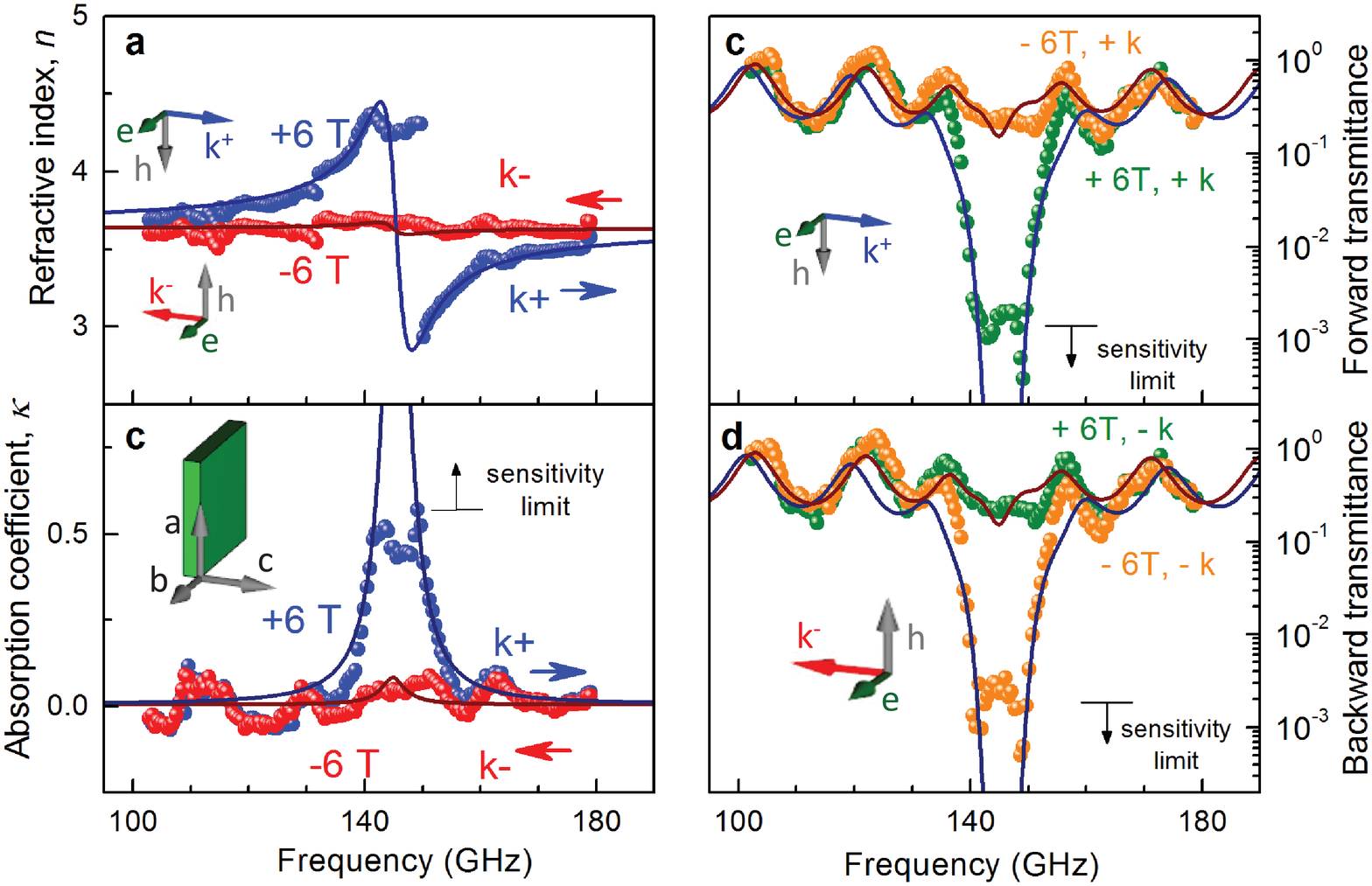}
\end{center}
\caption{Nonreciprocal transmission in \Sm.
Left panels:  \textbf{a} - Refractive index in \Sm~ for forward (blue) and backward (red) propagation of the millimeter-wave radiation. \textbf{b} - Absorption coefficient in \Sm~ for forward and backward propagation. In this experiments the inversion of the propagation is realized by the inversion of external magnetic field. Right panels: demonstration of the non-reciprocity  by explicit exchange of source and detector. \textbf{c} - Forward transmission for two opposite directions of external magnetic field. \textbf{d} - Backward transmission. The geometry of the experiment is given in the inset. Symbols - experiment, solid lines - fits assuming a Lorentzian form of the electromagnon\cite{kuzmenko_prb_2015}.} \label{fSmnonr}
\end{figure}

As was demonstrated in Fig.~\ref{fSmrot}\textbf{c,g}, in the geometry with external magnetic fields $\vect{H} \| b$--axis the electromagnon in \Sm~ is excited simultaneously via electric and magnetic channels: ${h}\|a$, ${e}\|b$. In this case the magnetoelectric coupling starts to distinguish between two possible propagation directions. This effect arises because electric polarization, $\vect{P}\|a$, is oriented perpendicular to the induced magnetization $\vect{M}\|\vect{H}\|b$. Consequently, \Sm~ gets a toroidal~\cite{fiebig_jpd_2005} moment, $\vect{T}=\vect{P}\times \vect{M}$, which may be parallel or antiparallel to the propagation direction~\cite{tokura_jmmm_2007}. The toroidal moment allows the existence of strong directional birefringence in \Sm.

The susceptibility matrices for this geometry can be written in a simplified form\cite{kuzmenko_prb_2015}:
\begin{equation}
\begin{array}{cc}
\hat{\chi}^m(\omega) =
\left( \begin{array}{ccc}
\chi_{xx}^m & 0 & \chi_{xz}^m \\
0 & \chi_{yy}^m & 0 \\
-\chi_{xz}^m & 0 & \chi_{xx}^m \\
\end{array} \right) & \hat{\chi}^{me}(\omega) =
\left( \begin{array}{ccc}
0 & \chi_{xy}^{me} & 0 \\
0 & 0 & 0 \\
0 & \chi_{zy}^{me} & 0 \\
\end{array} \right) \\[3em]
\hat{\chi}^{em}(\omega) = \left( \begin{array}{ccc}
0 & 0 & 0 \\
\chi_{xy}^{me} & 0 & -\chi_{zy}^{me} \\
0 & 0 & 0 \\
\end{array} \right) & \hat{\chi}^e(\omega) = \left( \begin{array}{ccc}
\chi_{xx}^e & 0 & 0 \\
0 & \chi_{yy}^e & 0 \\
0 & 0 & \chi_{zz}^e \\
\end{array} \right). \\
\end{array}
\label{susceptibilities}
\end{equation}

The set of basic equations to obtain the electromagnetic waves for the case of directional anisotropy (nonreciprocal transmission) is discussed by M. Mochizuki, DOI:10.1515/PSR.2019.0017. Here, we reproduce only the solutions of the Maxwell equations.
For a linearly polarized electromagnetic wave with $e \| y$, $\vect{h}\| x$  propagating along the $z$-direction the complex refractive indexes of the forward ($n^+$) and backward ($n^-$) solutions are different and they are given by:
\begin{equation}\label{eqnk}
  n^{\pm}= \sqrt{\tilde{\varepsilon}_y \tilde{\mu}_x} \pm \tilde{\alpha}_{xy} \ .
\end{equation}

Here $\tilde{\alpha}_{xy} = \chi_{xy}^{me} -\chi_{zy}^{me} \chi_{xz}^{m} / \mu_{z}$, $\tilde{\varepsilon}_y = \varepsilon_y + (\chi_{zy}^{me})^2/\mu_{z}$, and $\tilde{\mu}_x = \mu_x + (\chi_{xz}^{m})^2/\mu_{z}$ are the renormalized material parameters;
$\varepsilon_y = 1+ \chi_{yy}^e$ is the dielectric permittivity along the $y$-axis, $\mu_x= 1+ \chi_{xx}^m$ and $\mu_z= 1+ \chi_{zz}^m$ are the magnetic permeabilities along the $x$-axis and $z$-axis, respectively. The solution for the perpendicular polarization with $e \| x$, $\vect{h}\| y$ is trivial and can be written as $n^{\pm}= \sqrt{\varepsilon_x \mu_y}$. Importantly, two possible solutions for the propagating wave are linearly polarized. Therefore, no polarization rotation is expected for the geometry with $\vect{H}\|b$ and the two linear polarizations do not mix. This is in contrast to a related geometry with $\vect{H}\|a$ revealing optical activity which has been discussed in Sec.~\ref{chopt}. Figure \ref{fSmnonr} presents the experimental results on directional birefringence and directional dichroism in \Sm.

\begin{figure}[tbp]
\begin{center}
\includegraphics[width=0.86\linewidth, clip]{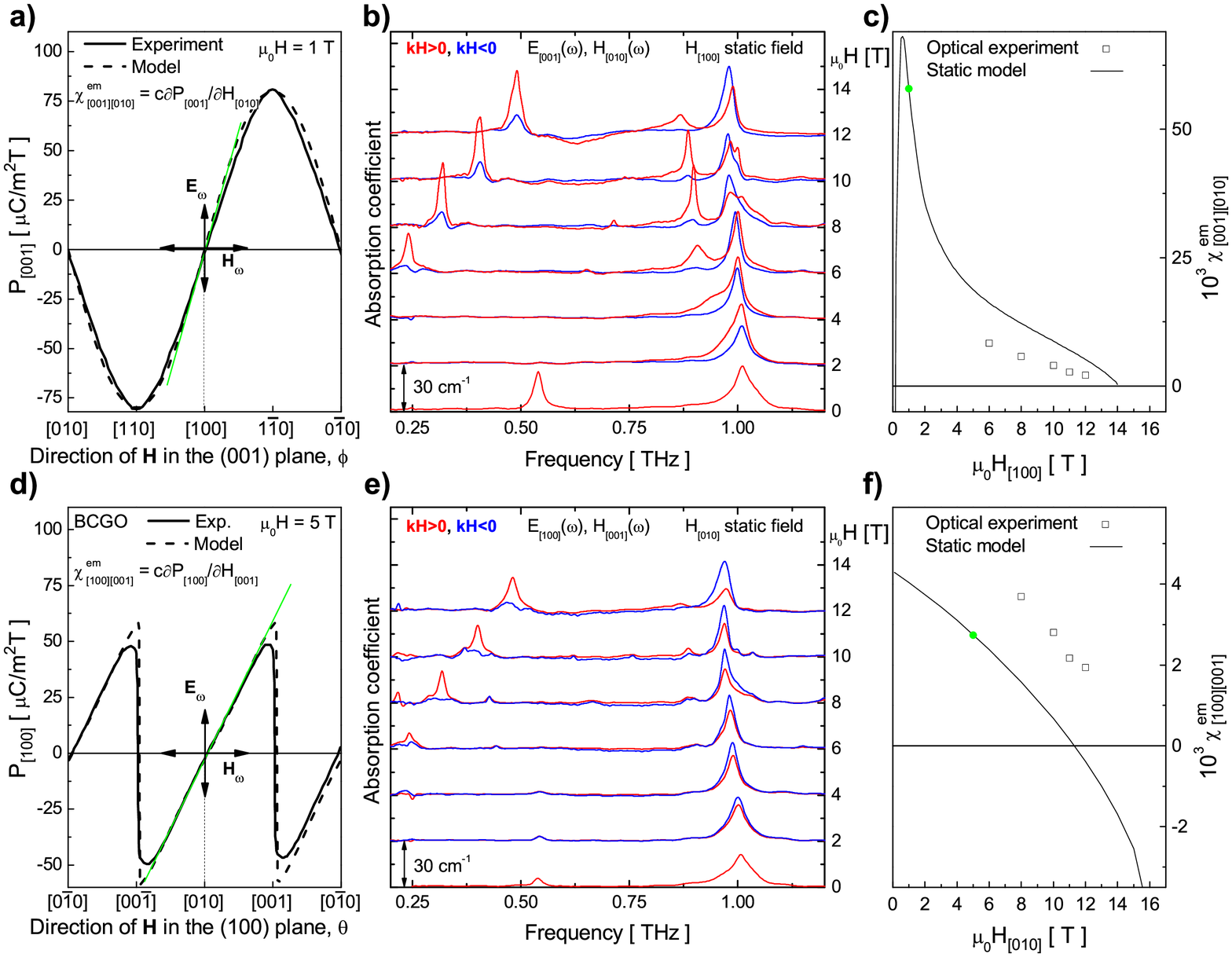}
\end{center}
\caption{Static and dynamic magnetoelectric properties of \Ba~(BCGO)).
Change of the electric polarization ($P$) when rotating the magnetic field ($H$) in the $(001)$ (a) and $(100)$  (d) crystal plane, respectively.
Experimental results\cite{murakawa_prl_2010} and model calculations using Eqs.~(\ref{Pa}) and (\ref{Pc}) are presented by solid and dashed line, respectively.
Here the magnetoelectric susceptibility is indicated by the slope of green lines. Arrows labelled by ${\bf H}^\omega$ and ${\bf E}^\omega$ show the magnetic and electric components of light in the analogous optical experiment.
The absorption measurements\cite{bordacs_nphys_2012,szaller_prb_2014} presented in (b)/(e) correspond to the static experiments in (a)/(d).
Propagation along/opposite to the magnetic field is indicated by red/blue lines. The spectra are vertically shifted proportionally to the magnetic field.
(c) and (f) show the field dependences of static magnetoelectric susceptibilities resulted from  static (lines)  and optical  experiments (symboles).
Green dots in (c) and (f) correspond to the slope of the green lines in (a) and (d), respectively. Reproduced from Ref.~[\onlinecite{szaller_prb_2014}].
} \label{fBanonr}
\end{figure}

\subsection{One-way transparency by magneto-chiral dichroism in multiferroic \Ba} \label{DA_BCGO}

According to the magnetoelectric sum rule in Eq.~(\ref{KhiME_DC}), the static
magnetoelectric coupling in Eqs.~(\ref{Pa}-\ref{Pc}) is a precursor of the
direction-dependent absorption ---directional dichroism--- appearing in the
low-frequency spin excitations of \Ba. Indeed, Ba$_2$CoGe$_2$O$_7$ was the
first compound exhibiting considerable directional dichroism in the THz
frequency range.\cite{kezsmarki_prl_2011} Here we focus on the one-way
transparency observed in the MChD
geometry\cite{bordacs_nphys_2012,kezsmarki_nc_2014,szaller_prb_2014}, when
the chirality of the system is induced by the magnetic field along the
$[100]$ or $[010]$ axes.

In Figs.~\ref{fBanonr}\textbf{b} and \ref{fBanonr}\textbf{e}, following Refs.\,[\onlinecite{bordacs_nphys_2012, szaller_prb_2014}], we review the MChD spectra of \Ba. The clear difference in the absorption of waves propagating parallel and antiparallel to the magnetic field is observable both for the quasi-ferromagnetic mode $(\omega\sim H)$ and for the exotic spin-streching excitation\cite{penc_prl_2012} $(\omega\approx 1\textrm{~THz})$. A reference experiment\cite{bordacs_nphys_2012} with magnetic field and light propagation along the $[110]$ direction showed no directional dichroism, corresponding to the restored $m_{[110]}^\prime$ and $m_{[1\overline{1}0]}$ mirror symmetries. Moreover, repeating the measurement\cite{kezsmarki_nc_2014} for $\vect{k}\parallel\vect{H}\parallel [010]$ resulted in a directional dichroism of opposite sign and equal magnitude as compared to the $\vect{k}\parallel\vect{H}\parallel [100]$ case, proving the magneto-chiral origin of the effect.

Using the sum rule in Eq.~(\ref{KhiME_DC}), the directional dichroism spectra
can be connected to the static magnetoelectric
experiments\cite{murakawa_prl_2010}. Fig.~\ref{fBanonr}\textbf{d} shows
$P_{[100]}(\theta)$, as reproduced from
Ref.\,[\onlinecite{murakawa_prl_2010}], where $\theta$ is the angle between
the magnetic field and the $[001]$ axis. In the first-order approximation,
the tilting of the magnetic field from the $[010]$ direction by
$\delta\theta$ can be considered as an additional transversal field
$\delta{\bf H}=(0,0,H\textrm{sin}\delta\theta)$, thus
\begin{eqnarray}
\chi^{em}_{[100][001]}\left(H_{[010]}=H\right) &=& c\at{\frac{\partial P_{[100]}}{\partial H_{[001]}}}{H_{[010]}=H}\approx \frac{c}{H} \at{\frac{\partial P_{[100]}}{\partial \theta}}{H_{[010]}=H}\nonumber\\
&\approx & \lim_{\delta\to 0+} \frac{2c}{\pi H} P_{[100]}\left( H_{[0\delta 1]}=H\right).
\end{eqnarray}
Here $\lim_{\delta\to 0+} P_{[100]}\left( H_{[0\delta 1]}=H\right)$ is equivalent to the amplitude of the $P_{[100]}\left( \theta\right)$ function. In the analogous optical experiment $\vect{H}\parallel [010]$, while the oscillating magnetic field is parallel to $\delta{\bf H}\parallel [001]$ and the electric field of light oscillates along $[100]$. The corresponding THz-frequency absorption spectra are shown for the two opposite propagation directions (red and blue lines) in Fig.~\ref{fBanonr}\textbf{e}, where the directional dichroism can be described by $\Delta\alpha(\omega)=\frac{4\omega}{c}\Im \chi^{em}_{[100][001]}(\omega)$.

Similarly, the  $P_{[001]}(\phi)$ curve in Fig.~\ref{fBanonr}\textbf{a}, as taken from Ref.\,[\onlinecite{murakawa_prl_2010}], can be used to calculate $\chi^{em}_{[001][010]}$ in the static limit:
\begin{eqnarray}
\chi^{em}_{[001][010]}\left(H_{[100]}=H\right) &=& c\at{\frac{\partial P_{[001]}}{\partial H_{[010]}}}{H_{[100]}=H} \approx \frac{c}{H} \at{\frac{\partial P_{[001]}}{\partial \phi}}{H_{[100]}=H} \nonumber\\
&\approx & \frac{2c}{H} P_{[001]}\left( H_{[110]}=H\right)
\end{eqnarray}
for $\vect{H}\parallel [100]$, where $P_{[001]}\left( H_{[110]}=H\right)$ is the amplitude of $P_{[001]}\left(\phi\right)$. In the optical range, $\vect{H}\parallel [100]$, oscillating magnetic and electric fields along $[010]$ and  $[001]$ result in directional dichroism corresponding to $\chi^{em}_{[001][010]}$.

Figs.~\ref{fBanonr}\textbf{c} and \ref{fBanonr}\textbf{f} shows the comparison between the field dependencies of magnetoelectric tensor elements obtained from the static and optical experiments. While the order of magnitude and the tendency in the field dependence is similar, there is a significant difference in the numerical values, which can be attributed to the different experimental conditions (temperature, sample growth, uncertainty in the geometrical parameters of the sample, etc.)\cite{szaller_prb_2014}.

\section{Summary}

The appearance of the magnetoelectric term in the Maxwell equations leads to novel possibilities to control the propagation of light. The new effects are present in several multiferroics because of internal coupling between electric and magnetic order. The electromagnons, new electromagnetic excitations in multiferroics, extend the magnetoelectric coupling to the dynamic regime. Due to resonance-like character of the electromagnons, new optical effects may be especially strong close to the resonance frequency.

In this contribution we considered two major optical effects due to dynamic magnetoelectric coupling: optical activity and directional anisotropy.
As has been demonstrated in Section~\ref{s_OA}, in case of multiferroics the optical activity may be reduced to magnetoelectric susceptibility and in a simple approximation is even proportional to it. As two characteristic examples, the polarization rotation of \Sm~ and \Ba~ have been considered, both demonstrating large effects close to the electromagnon frequencies.

Depending on the magnetic symmetry, the existence of the magnetoelectric term may lead to another important effect, the directional anisotropy. Besides symmetry considerations, direct solution of the Maxwell equations allows us to look for a difference in forward and backward propagation. Directional anisotropy arises when two related eigenmodes, corresponding to opposite propagation directions, reveal different propagation constants. Here we demonstrated that both, \Sm\, and \Ba, reveal strong directional anisotropy under certain experimental conditions.

\section{Outlook}

In spite of the richness of optical effects reported in multiferroic materials, several questions remain open. One example of an unexplored optical effect is given by the gyrotropic birefringence, i.e. kind of polarization rotation that reverses sign both, under time inversion and under space inversion. Presently, a classification of possible optical effects and their connection to specific terms in the magnetoelectric susceptibility is still missing. This classification should also include the symmetry analysis of the magnetoelectric coefficients and suggest magnetic structures for possible realizations.

Both, optical activity and asymmetric transmission described above are sensitive to the polarization state of the radiation. That is, these effects can be observed basically for one solution of the Maxwell equations only. The other solution remains in several cases unaffected. Therefore, especially for the case of asymmetric transmission it is highly desired to look for magnetoelectric materials with polarization-independent directional anisotropy.

Ordered magnetic structures that are necessary for the optical effect exist mainly at rather low temperatures, or at high magnetic fields\cite{Viirok_prb_2019}. This is a disadvantage with respect to possible application of multiferroics in optical communication. In the static regime, present success to push the phase transitions in multiferroics above the room temperature is remarkable. In case of dynamical properties, except for the weak directional dichroism of BiFeO$_3$\cite{kezsmarki_prl_2015}, the room temperature effects still have to be found.

\section*{Acknowledgements}
This work was supported by the
Russian Science Foundation (16-12-10531) and by
the Austrian Science Funds (W 1243, I 2816-N27, I 1648-N27).

\bibliographystyle{unsrt}
\bibliography{literature}

\end{document}